\documentclass[aps,prb,twocolumn,showpacs,floatfix]{revtex4}
\usepackage{graphicx}

\begin{document}

\newcommand{\zn}{ZnV$_2$O$_4\,\,$}
\newcommand{\lc}{\lowercase}

\title{Orbital order in ZnV$_2$O$_4$}

\author{Tulika Maitra, Roser Valent{\'\i}}

\affiliation{ 
Institut f{\"u}r Theoretische Physik, J. W. Goethe-Universit{\"a}t,
Max-von-Laue-Str. 1, 60438 
Frankfurt, Germany}

\date{\today}

\begin{abstract}

In view of recent controversy regarding the orbital order
in the  frustrated spinel \zn,   we analyze the orbital
and magnetic groundstate of this system  within an  {\it ab initio}
density functional theory approach.   
 While LDA+U  calculations in the presence of a cooperative Jahn-Teller
 distortion
stabilize an A-type staggered orbital order, the consideration of
relativistic spin-orbit effects unquenches the orbital moment and leads to a
uniform orbital
order
with a 
 net magnetic moment close to 
the experimental one.  Our results 
show that  {\it ab initio} calculations are able to resolve the existing discrepancies
in previous theories and that it is the spin-orbit coupling alongwith  electronic correlations 
 which play a significant role in determining the
  orbital structure in these materials.

\footnotesize{}
 
\end{abstract}
\maketitle

Correlated electronic systems involving transition metal oxides took
the centrestage of condensed matter physics research for the last three decades 
because of their intriguing, often non-intuitive properties. Transition
metal spinel oxides with an additional complexity of a geometrically frustated lattice provide 
an exciting ground for the study of several competing interactions among 
spin, orbital and lattice degrees of freedom\cite{suzuki,lee}. 
Besides being of fundamental interest, spinels have been
also proposed for 
 spintronics applications\cite{lueders,hu}.
   In the present work on \zn we investigate the effect
of competing spin, orbital and lattice degrees of freedom and
show that density functional calculations provide
an adequate and realistic ground  to establish the 
dominant mechanism driving the orbital order in vanadium spinels.
 
 The orbital order in
ZnV$_2$O$_4$ as well as in other vanadium spinels such as MgV$_2$O$_4$ and CdV$_2$O$_4$ is presently a subject of considerable debate \cite{radaelli}.
In order to understand the behavior of these compounds,
 various groups\cite{motome,tchernyshyov,matteo} have proposed 
 alternative 
microscopic mechanisms which predict
different orbital patterns. 
The ongoing debate has its origin in the complex nature
of these systems with competing spin, orbital and lattice
degrees of freedom. These systems
 have V$^{3+}$ ions
   in a spin 1 state characterized by double occupancy
 of the triply degenerate $t_{2g}$  ($d_{xy}$, $d_{xz}$, $d_{yz}$)
orbitals. These partially filled $t_{2g}$ orbitals
 leave the  orbital degrees of freedom unfrozen opening up the possibility
of orbital order.
Moreover, the V-sites in the cubic spinel structure form a pyrochlore lattice,
which 
gives rise to  frustrated antiferromagnetic interactions among these sites\cite{pisani06}.
In \zn the 
interplay
of all these degrees of freedom leads to
two successive phase transitions which involve structural, orbital
and magnetic changes. At T$_S$= 51 K, \zn undergoes
a structural phase transition where the symmetry is
 lowered from cubic to tetragonal with a compression of the
 VO$_6$ octahedron along the $c$ axis\cite{reehuis}
and the system possibly orbital orders.
 The structural transition also lifts the geometrical frustration of the cubic phase
 making a way for the second transition at T$_N$= 40 K which
 is of magnetic nature and the system orders antiferromagnetically
 \cite{reehuis,ueda}.

\begin{figure}
\includegraphics[width=8cm]{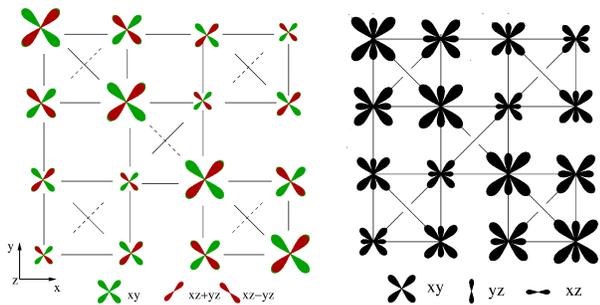}
\caption{(Color online) Schematic diagram showing the orbital order obtained
  within LDA+U in this work (left panel) and that proposed in Ref. \onlinecite{motome} 
  (right panel).}
\label{schmfig}
\end{figure}

 While the antiferromagnetic structure in \zn at T $<$ T$_N$
has been unambiguously
determined by neutron scattering experiments
at low temperatures\cite{niziol_73}, the nature
of the orbital groundstate in the whole temperature 
range T $<$ T$_S$
is still unclear.  
 Tsunetsugu and Motome\cite{motome} proposed that
the Coulomb and exchange interaction between the magnetic ions
as expressed in the Kugel-Khomskii Hamiltonian 
and the coupling to the Jahn-Teller lattice distortion are the 
driving mechanisms of the consecutive phase transitions. 
The groundstate orbital order predicted by their model  has an 
alternating singly occupied $d_{xz}$ and $d_{yz}$ orbital along the $c$ direction together with
a singly occupied $d_{xy}$ orbital in all V sites (see Fig.\ \ref{schmfig}, 
right panel). They showed
that this type of orbital order is consistent with the observed
antiferromagnetic order at low temperatures\cite{niziol_73,ueda}  as well
 as with the compressed VO$_6$ octahedron in the tetragonal phase\cite{reehuis}.
But this orbital order  breaks the mirror reflections in the
planes (110) and (1$\bar{1}$0) and the diamond glides  
 in the planes (100) and (010) and hence is found to be  
inconsistent with the spatial symmetry $I4_1/amd$ predicted by
 X-ray scattering 
experiments on polycrystalline samples \cite{reehuis,nishiguchi}. 
 Recent inelastic neutron scattering data on a 
single crystal sample\cite{lee} seem to suggest though a breaking of the glide
symmetry. 
  These results have nevertheless not been further
corroborated by other measurements.

Tchernychyov\cite{tchernyshyov}, in a separate theoretical work,
 considering the fact that V$^{3+}$ has a non-negligible orbital
moment,  proposed 
a single-ion 
model where  
the relativistic spin-orbit coupling plays the dominant role in driving the orbital 
order in the  tetragonal phase.    
 Including the Jahn-Teller effect as a perturbation in his model, 
Tchernyshyov predicted a uniform (ferro) orbital order with a complex orbital 
state 
$d_{xz}$$\pm$i$d_{yz}$ at 
each V-site. This orbital order is consistent with the symmetry $I4_1/amd$ 
since it preserves both mirror and glide symmetries and hence is at odds
with the prediction of Tsunetsugu {\it et al.}\cite{motome}.  
 Di Matteo {\it et al.}\cite{matteo} in a subsequent work
 proposed a third alternative model
where both 
the relativistic spin-orbit coupling and the 
Kugel-Khomskii model of spin-orbital superexchange 
are treated on an 
equal footing. They observed that for intermediate spin-orbit
coupling strength the groundstate of the system shows a uniform
orbital order compatible with Tchernychyov's prediction and with the
symmetry $I4_1/amd$.

In view of the above mentioned disagreements among the various theoretical
models, we analyze in this work the
orbital and magnetic groundstate of \zn in the tetragonal
 phase within an 
  {\it ab initio} density functional theory (DFT)
approach.  Our calculations without spin-orbit coupling
 indicate the presence of a staggered antiferro-orbital
order compatible with the spatial symmetry of the tetragonal phase $I4_1/amd$ and
we observe that correlation effects are essential for the description of
this state.
These results agree with the prediction of Tsunetsugu and Motome\cite{motome} about
the existence of a staggered orbital order except that
 we obtain a pattern which,
 contrary to Ref. \onlinecite{motome}, is compatible
 with the underlying $I4_1/amd$  symmetry (see Fig. \ref{schmfig}). 
Inclusion of relativistic spin-orbit effects in our DFT
calculations changes though this
picture considerably.    
   For moderate to large values
of the onsite electronic correlation U, a uniform orbital order
analogous to the cubic phase\cite{Anisimov99,Eyert} -where the
orbitals are tilted due to the trigonal distortion of the structure-
is stabilized.  Also,  a large orbital moment antiparallel to
the spin moment is obtained
in the presence of strong onsite correlation which 
gives rise to a net magnetic moment in close agreement with the 
experimentally observed one.
 Finally, we confirm that the antiferromagnetic order 
 observed in neutron 
scattering experiments\cite{reehuis,ueda}  minimizes the {\it ab
  initio}
DFT calculations.

 Our  DFT calculations were performed
within  the local spin density approximation (LSDA)  
and the  LSDA+U\cite{Anisimov} approach.     
 The spin-orbit coupling was treated using a scalar-relativistic basis
and the second variational method\cite{koelling77} (LSDA+U+SO). 
 All calculations were done
with  the 
 full-potential linearized augmented plane-wave code 
WIEN2k \cite{wien2k}.
The atomic sphere radii were chosen to be 1.96, 1.99 and 1.78 a.u. for
Zn, V and O respectively and 40 ${\bf k}$ points mesh in the
irreducible wedge was considered for Brillouin zone integrations 
We  used the 
structure data  measured in Ref.  \onlinecite{reehuis}. 

\begin{figure}
\includegraphics[width=7cm]{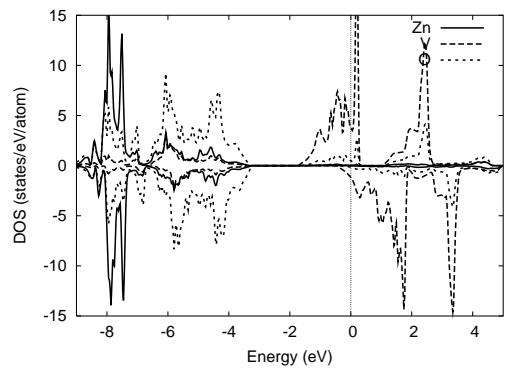}
\caption{Total density of states for Zn, V, and 
O in both spin up and down channel in LSDA.
 }
\label{dos-lsda}
\end{figure}

In Fig. \ref{dos-lsda} we present the
spin polarized density of states (DOS) of Zn, V and O within the
LSDA for  both spin directions. 
The V t$_{2g}$ up spin 
states are partially occupied defining a  metallic solution within the
LSDA approximation.
The calculated 
  magnetic moment per vanadium is  1.48$\mu_B$.
 The above results show that  V$^{3+}$ is in a high spin ($S=1$) state implying
 strong Hund's coupling (J$_H$) strength.
Due to the $c$-axis compression of
the VO$_6$ octahedra,  there is a small 
but finite splitting of the $d_{xy}$-level from the doubly degenerate
$d_{xz}$/$d_{yz}$ level (not shown here).

 \zn is  a Mott insulator, the inclusion of correlation effects
are expected to improve on the incorrect metallic solution given by the LSDA 
approach. It is also well known\cite{kugel,liechtenstein} that in  Mott
 insulators with orbital degeneracies as in our case, orbital order
plays a crucial role in driving any collective Jahn-Teller distortion present
in the system.  
Calculations within the LSDA+U approach 
-where orbital dependent potentials are included-  should then be
more appropriate to describe the gap and the possible orbital order
in these systems.

\begin{figure}
\includegraphics[width=7cm]{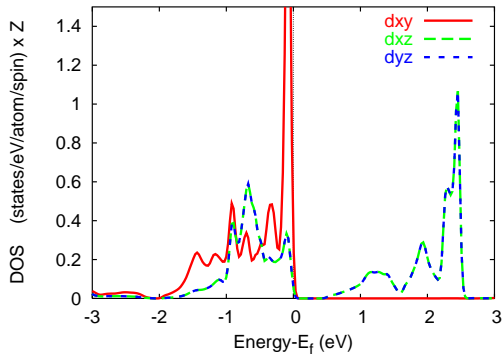}
\caption{Projected V $t_{2g}$ DOS 
 in the spin up channel within the LSDA+U approach.   }
\label{dos-ldau}
\end{figure}

\begin{figure}
\includegraphics[width=8.0cm]{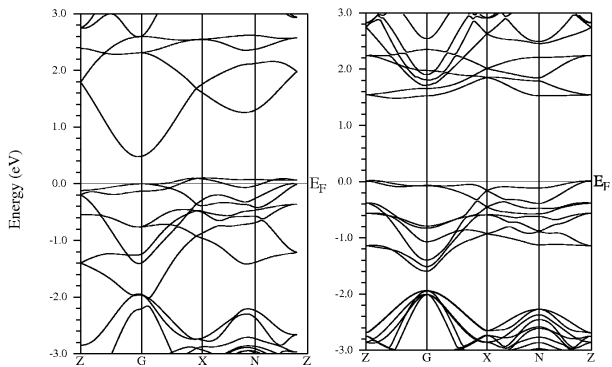}
\caption{Bandstructure in the energy interval
$(-3, 3) eV$  within LSDA+U (spin up channel)
(left) and LSDA+U+SO (right). The bands are shown along the
path $Z(0,0,2\pi/c) - G(0,0,0) - X(\pi/a,\pi/a,0) - N(\pi/a,0,\pi/c) - Z$. Note the narrowing of
the broad V $d$  unoccupied band ( energy between 0.4 eV and 1.8 eV)
when considering SO.}
\label{bands}
\end{figure}

In Fig. \ref{dos-ldau}  we present
  the projected DOS of the V  
$t_{2g}$-states in the up  spin channel  calculated within
 the LSDA+U  with $U$=5 eV and
$J$= 0.9 eV 
\cite{mizokawa}.
 Note that the
  d$_{xy}$ up band now gets completely occupied with one 
electron  and the partially filled 
doubly degenerate $d_{xz}$/$d_{yz}$
band observed in LSDA is split into two opening up a gap of
0.4 eV. The second electron 
now occupies the lower energy up
band which  has  both $d_{xz}$ and $d_{yz}$ character. 
In Fig.\ \ref{bands} (left panel) we show the corresponding bandstructure.
The orbital ground state described below is found to be
stable with $U$ values in the range 4 - 6 eV.

\begin{figure}
\includegraphics[width=6.5cm]{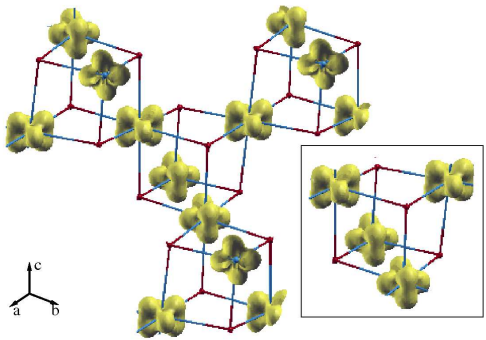}
\caption{(Color online) Three dimensional electron density plots showing the 
staggered orbital order in this system. The inset shows the occupation of 
vanadium t$_{2g}$ orbitals in an elementary V$_4$O$_4$ cube. The isovalue used 
here is 0.6e/A$^3$.}
\label{el-den1}
\end{figure}

  We present in Fig.\  \ref{el-den1} a three-dimensional
 plot of the spin density of
\zn calculated within LSDA+U in the range of energies between -2 eV and
 the Fermi level. This range of energies includes V $t_{2g}$ and O-$p$ states. 
 A close inspection of Fig.  \ref{el-den1}  reveals that 
all V ions in the tetragonal
spinel structure  have one filled t$_{2g}$ orbital  in the 
$ab$-plane ( $d_{xy}$   in the
 reference frame  defined with the $a$, $b$ axes directed along the basal
 V-O$_b$ bonds and $c$ axis directed along the apical V-O$_a$ bond).
This feature is already observed in the projected DOS in Fig. \ref{dos-ldau}.
  The second vanadium electron occupies a $t_{2g}$ orbital with
 symmetry  $d_{xz}+d_{yz}$ and
$d_{xz}-d_{yz}$ in  alternate $ab$-planes respectively along the
$c$-direction (please note  the four lobes in planes perpendicular
 to $ab$  in Fig. \ref{el-den1}).  
The V$_4$O$_4$ cube (Fig. \ref{el-den1} (inset))  shows two V  with the second t$_{2g}$ electron in 
 a $d_{xz}+d_{yz}$ symmetry and two V with the second $t_{2g}$ electron in a
 $d_{xz}-d_{yz}$ 
symmetry.

This orbital structure is therefore antiferro-orbitally ordered
along $c$  (A-type) and ferro-orbitally ordered in the $ab$ plane.
 In comparison, in the high 
temperature cubic phase the t$_{2g}$ orbitals are  split into singlet   $a_{1g}$ and  doublet 
$e_g$$^{\prime}$ due to a small trigonal distortion
present in the system.  The higher energy doublet is equally occupied at 
each V-site implying the absence of any orbital order unlike the tetragonal
phase. Note that the trigonal distortion is also present in the low temperature
tetragonal phase alongwith the tetragonal distortion. Here we discuss our results
in the ($d_{xy}$, $d_{yz}$, $d_{xz}$)-basis instead of $a_{1g}$ and $e_g$$^{\prime}$.

While the A-type orbital ordering observed by Motome and Tsunetsugu 
\cite{motome} for \zn with alternately occupied $d_{xz}$ and $d_{yz}$ orbital
along the $c$-direction (Fig. \ref{schmfig} right panel) is found to be
inconsistent with the underlying space group symmetry of the crystal 
($I4_1/amd$), 
 the  orbital order described in the present work is compatible with
 the crystal symmetry as it is obtained from a self-
consistent field calculation on the  crystal 
structure described by the $I4_1/amd$  spatial group (Fig. \ref{schmfig} left 
panel).

The orbital order described above is 
driven by the combination of the staggered distortion present in the experimental
lattice structure with correlation effects.
First of all, the cubic to tetragonal transition with a compression of the 
VO$_6$ 
octahedron along the $c$-axis splits the triply degenerate t$_{2g}$ states
into a lower energy  $d_{xy}$-orbital and higher energy doublet which are linear
combinations of $d_{xz}$ and $d_{yz}$ orbitals
 (i.e. $d_{xz} \pm d_{yz}$), then the degeneracy in the
higher energy doublet is  lifted at each vanadium site due to a
co-operative Jahn-Teller (JT)-like distortion (combination of trigonal 
distortion of the V sites) present in the experimental lattice 
structure which makes the O-V-O angles to alternate between
85$^{\circ}$ and 95$^{\circ}$ in a staggered fashion along $c$-axis. Hence  
the faces of V$_4$O$_4$ cubes in the $ab$-plane alternately expand or
compress along the $c$-axis (see Fig. \ref{el-den1}). LDA calculations, due
to its known inabilities to describe this type of electronically driven 
cooperative JT distortion \cite{liechtenstein,kugel} fail 
to lift the degeneracy of the higher energy doublet and  both $d_{xz} \pm d_{yz}$ orbitals 
are partially filled at each 
V-site. Consideration of $U$ completely lifts this degeneracy and an orbital
ordering 
is obtained
as shown in Fig. \ref{el-den1}.

In the next step of our calculations
we  perform a 
relativistic calculation including the spin-orbit interaction (LSDA+U+SO) in
the second variational method. 
The magnetization direction is set along the $c$-axis as the experimentally
observed spontaneous magnetic moment points to this direction and also this
is the high symmetry direction of the tetragonal crystal structure. 
\begin{figure}
\includegraphics[width=6.5cm]{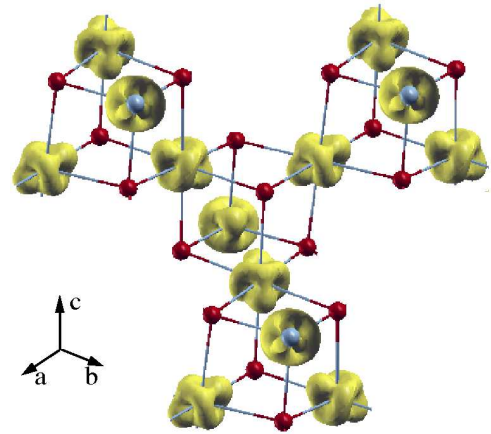}
\caption{(Color online) Three dimensional electron density plots showing the 
uniform orbital order within LSDA+U+SO.}
\label{soordering}
\end{figure}
Within the LSDA+U+SO approach a uniform orbital order (see
   Fig.\ \ref{soordering}) is favoured in contrast to 
the 
staggered A-type order found within the LSDA+U only approach. The 
orbital state at each site 
results from the mixture of the $d_{xz}+id_{yz}$ and $d_{xz}-id_{yz}$ 
orbitals that were found to be staggered at alternate sites within the LSDA+U.
Due to the partial occupation of both  $d_{xz}+id_{yz}$ and  $d_{xz}-id_{yz}$ orbitals,
the system has  a finite l$_z$= 0.75 $\mu_B$
moment.   Since this
orbital magnetic moment is antiferromagnetically coupled to the spin only
magnetic moment (1.69 $\mu_B$) we obtain a total magnetic moment 
of 0.94$\mu_B$ per vanadium ion. We note that the good agreement 
with the experimentally observed magnetic moment (0.63 $\mu_B$) \cite{reehuis} 
implies
an important contribution of the orbital degrees of freedom in the present 
system. In fact, LSDA+SO only calculation provides a small orbital 
moment of about 0.04 $\mu_B$ due to the inability of LSDA to treat orbital
polarization effects~\cite{novak}. Including atomic orbital correlations on top of 
LSDA+SO scheme through the Hubbard $U$ allows for an appropriate description 
of the atomic orbital state.

In order to understand the effects of the spin-orbit coupling, we compare in 
Fig.\ \ref{bands}  the  bandstructure within LSDA+U (left panel) and LSDA+U+SO (right panel). In case of LSDA+U the bands are shown in the spin up channel 
because the bandgap is formed in this channel whereas the spin down bands are about 1 eV above the Fermi level.    
 The significant increase in the bandgap within LSDA+U+SO approach 
(1.0 eV )
can be understood by the fact that the small but finite orbital moment($\sim
0.04 \mu_B$) obtained within LSDA+SO is driven to about 0.75 $\mu_B$ by the
orbital polarization effects of $U$. 
With this large orbital moment, the spin-orbit interaction term in the Hamiltonian
pushes the unoccupied $d$-bands above the Fermi level further up in energy. We 
also observe a decrease in the bandwidth of    
these bands compared to the case of LSDA+U and thereby a substantial increase in
the bandgap.

An experimental measurement of the band gap in this system could
further establish the role of spin-orbit interaction and atomic orbital 
correlation and thereby corroborate the type of orbital ordering.
Our calculations show that this (uniform) orbital ordering persists in 
the antiferromagnetic phase  observed
in INS\cite{reehuis,ueda}.

In summary, we have investigated the orbital order in 3$d^2$ magnetically 
frustrated systems by performing 
{\it ab-initio} DFT calculations on \zn and observe that although strong onsite 
electronic 
correlation effects in the presence of co-operative JT distortion stabilize
an A-type staggered orbital order, the consideration of relativistic spin-orbit 
effects  unquenches the orbital moment and leads to a uniform orbital order state with a large
orbital moment.
This orbital moment couples 
antiferromagnetically to the spin moment of V with
a net magnetic moment close to the experimentally observed one.
This study shows that for a realistic description of the orbital order in
\zn it is the interplay between electron correlation, spin-orbit coupling  
and co-operative JT distortions what  drives the orbital order. 
Our results help to resolve the differences among all the previous theories
and present a consensus picture of the orbital order in \zn compatible with the experimental 
findings.  
Furthermore, this work  shows direct
evidence from {\it ab-initio} calculations of the importance of spin-orbit effects in
 3$d^2$ systems.

\noindent{\bf Acknowledgement}
The authors would like to thank L. Pisani, O. Tchernyshyov, G. Jackeli, N.B. Perkins, T. Saha-Dasgupta,
H. Tsunetsugu and A. Yaresko
for useful discussions. R.V. thanks the DFG for financial support.


\end{document}